\documentclass[a4paper,12pt,english]{article}
\usepackage{amsfonts,bm,amssymb,euscript,array,babel,cite,amsmath,amsthm,amscd,graphicx}
\usepackage[all]{xy}



\newcommand{\be}{\begin{equation}} \newcommand{\ee}{\end{equation}}
\newcommand{\beq}{\begin{equation}} \newcommand{\eeq}{\end{equation}}
\newcommand{\beqa}{\begin{eqnarray}}
\newcommand{\eeqa}{\end{eqnarray}} 
\def\nn{\nonumber} \def\bea{\begin{eqnarray}} \def\eea{\end{eqnarray}}

\newcommand{\barr}{\begin{array}}
\newcommand{\earr}{\end{array}}

 \def\L{\Lambda}

 \def\one{\mbox{1 \kern-.59em {\rm l}}}

\def\bit{\begin{itemize}} \def\eit{\end{itemize}} 
\def\({\left(} \def\){\right)}    

 \allowdisplaybreaks[3]
\begin{document}

\renewcommand{\title}[1]{\vspace{10mm}\noindent{\Large{\bf
#1}}\vspace{8mm}} \newcommand{\authors}[1]{\noindent{\large
#1}\vspace{5mm}} \newcommand{\address}[1]{{\itshape #1\vspace{2mm}}}

\parindent=0cm

\begin{titlepage}

\begin{center}

\title{\Large Axiology}
\vskip 3mm

\authors{Athanasios {Chatzistavrakidis${}^{*,}${\footnote{than@th.physik.uni-bonn.de}}},
Encieh {Erfani${}^{*,}${\footnote{erfani@th.physik.uni-bonn.de}}},\\[1ex]
Hans Peter {Nilles${}^{*,}${\footnote{nilles@th.physik.uni-bonn.de}}}, 
Ivonne {Zavala${}^{\dagger,}${\footnote{e.i.zavala@rug.nl}}}}

\vskip 3mm
\address{${}^*${\it Bethe Center for Theoretical Physics and Physikalisches Institut, University of Bonn \\ Nussallee 12, D-53115 Bonn, Germany}}

\address{${}^\dagger${\it Center for Theoretical Physics, University of Groningen \\ Nijenborgh 4, 9747 AG Groningen, The Netherlands}}
\vskip 1.6cm

\textbf{Abstract}

\end{center}
Axions might play a crucial role for the solution of the strong CP problem and explanation of cold dark matter in the universe. In addition they may find applications in the formulation of inflationary models for the early universe and can serve as candidates for quintessence. We show that all these phenomena can be described within a single framework
exhibiting a specific pattern of mass scales: the axionic see-saw. We also discuss the role of supersymmetry (susy) in this axionic system in two specific examples: weak scale susy in the (multi) TeV range and tele-susy with a breakdown scale coinciding with the decay constant of the QCD axion: $f_a\sim 10^{11}-10^{12}$ GeV.

\vskip 3mm

\begin{minipage}{14cm}
\end{minipage}

\end{titlepage}

\tableofcontents

\section{Introduction}

Axions are well motivated candidates for physics Beyond the Standard Model (BSM) of particle physics \cite{Kim1}. Most notably they have played an important role in the discussion
of three specific topics (the three ``useful'' axions):

\begin{itemize}

\item the solution of the strong CP problem \cite{PQ,Weinberg, Wilczek, Dine:1981rt, Kim:1979if},

\item the formulation of natural inflation \cite{Natural, NaturalWMAP},

\item candidates for quintessence \cite{ChoiQA, KimQA ,KN1, KN2}.

\end{itemize}

Moreover, the (so called invisible) QCD axion that has been postulated in the context of the strong CP problem provides a perfect candidate for Cold Dark Matter (CDM) \cite{Kim1, Abbott:1982af, Preskill:1982cy, Dine:1982ah}. This constitutes an alternative to the WIMP paradigm for CDM in other approaches to  physics BSM, such as  supersymmetry (susy),  which also address the hierarchy problem of the weak scale compared to the Planck scale. Susy solves the hierarchy problem  by postulating new supersymmetric partners at a scale not too far above the weak scale. Experiments at the LHC are currently exploring this energy range and might soon give a conclusive answer. An even more severe hierarchy problem is related to the small size of the vacuum energy (or the cosmological constant) and this does not have an obvious solution in the framework of supersymmetry. Thus we might
need alternatives to understand the hierarchy problems (or fine tuning problems) of nature, such as {\it e.g.~}anthropic reasonings \cite{Susskind}.

Axion scenarios are such that do not add anything new to the solution of the hierarchy problem. Here we consider axions and supersymmetry as two independent approaches to physics BSM. It is known that they can coexist in meaningful ways \cite{Nilles:1981py}, but they do not necessarily need each other. In the present paper we shall analyze this coexistence and discuss possible relations between the mass scales of these theories, as {\it e.g.~}the scale of supersymmetry breakdown and the axion decay constant.

All attempts of physics BSM would ultimately need an ultraviolet completion which we assume to be string theory. Axionic particles are abundant in string theory \cite{SW, axiverse} and the axions mentioned above can be integrated in the scheme. Generically the axion decay constant $f_a$ is expected to be of order of the string scale $M_{\rm{string}}$. Axions are perturbatively massless and receive masses via nonperturbative effects. Depending on the size of these effects we can have a variety of mass scales, as {\it e.g.~}from $10^{13}$ GeV for the inflationary axion down to $10^{-32}$ eV for the quintessential axion.

Apart from the well motivated axions mentioned above there could be a plethora of other axion-like particles \cite{axiverse, Arias:2012mb} with various applications in particle physics and cosmology. We concentrate here on the three ``useful'' axions and analyze whether they can be incorporated in a single scheme. As we said earlier, generically one would expect the axion decay constants to be of order of the string scale; however, the scale of a successful QCD axion is found to be hierarchically smaller: $f_a\sim 10^{9}-10^{12}$ GeV \cite{KC} (partially motivated through considerations concerning the CDM abundance of the universe). Thus string theory needs a mechanism to lower the scale of the QCD axion.

This brings us back to supersymmetry. String theories seem to need some amount of supersymmetry for internal consistency and the absence of tachyons. We do not know the scale of supersymmetry breakdown, but we expect it to be sufficiently small compared to the string scale, supported by explicit string theory constructions towards the Minimal Supersymmetric Standard Model (MSSM) \cite{Lebedev:2006kn}. Of course, this is all a question of experimental verification and we hope that experiments at the LHC will provide useful hints for the nature of physics BSM. Unfortunately, no sign of new physics has been seen there, except for the recently announced evidence for the existence of the Standard Model (SM) Higgs boson at approximately 125 GeV. Let us suppose that the Higgs mass is in this range. For the MSSM this mass scale is rather high, indicating a high scale of susy breakdown at the multi-TeV range \cite{Krippendorf:2012ir}. On the contrary, for the SM without supersymmetry this Higgs mass is rather low. Evolution of the Higgs self coupling $\lambda$ would lead to negative values at scales small compared to the Planck scale \cite{Giudice:2011cg, Degrassi:2012ry}. This would indicate a vacuum instability that would require new physics BSM ({\it e.g.~}supersymmetry) at this scale. For a Higgs mass of $125$ GeV the mass range of instability is $10^{8}-10^{15}$ GeV \cite{Degrassi:2012ry} and coincides with the scale of the QCD axion. Is this an accident? Supersymmetry at this scale could be a reason for vanishing $\lambda$ via a shift symmetry as noted in \cite{Hebecker:2012qp},
that is a property of many successful model constructions in (heterotic) string theory \cite{Lebedev:2006kn}.

In the present paper we speculate that this is an indication for the presence of a remote supersymmetry (tele-susy) somewhere in the range between multi-TeV and the axion scale.
Many axion properties are independent of the scale of supersymmetry breakdown and meaningful axion models can be constructed within this range.

The outline of the paper is as follows. In section~2 we summarize the properties of the three well motivated axions in detail. We then present a discussion of axion potentials in the multiaxion case to obtain a unified scheme in section~3. Such a scheme requires the consideration of at least four axions, as will be discussed in section~4. Section~5 is devoted to the interpretation of the unified scheme and its possible connection to the scale of supersymmetry breakdown. Conclusions and outlook will follow in section~6.

\section{Axion properties}

Our present understanding of the cosmological evolution of the universe is best described by the $\Lambda$CDM model, complemented with the inflationary paradigm. The $\Lambda$CDM model accounts for the observed present day accelerated expansion in terms of a tiny cosmological constant $\Lambda$, as well as the observed existence of CDM in the universe. Additionally, inflation provides the seeds for the formation of the large scale structures  that we observe today.
However, CDM, inflation and present day acceleration, still demand a compelling explanation. Is it possible to find a single origin for these three crucial ingredients of the most successful cosmological model known to date? In this note we will argue in favor of a common origin for these ingredients in terms of axion-like particles.

The existence of axions  was first postulated to solve the strong CP problem of QCD \cite{PQ, Weinberg, Wilczek}. They are Pseudo Nambu-Goldstone Bosons (PNGBs), associated with the spontaneous breaking of a global ``Peccei-Quinn'' $U(1)_{\rm{PQ}}$ symmetry at scale $f_{a}$, which acquire a mass via QCD anomaly (instanton) effects. The decay constant $f_a$ and mass of the QCD axion are constrained by observations, leaving only a parameter window for the QCD invisible axion: $10^{9}\,{\rm GeV}\lesssim f_a\lesssim 10^{12}\,{\rm GeV}$ and $10^{-6}\,{\rm eV}\lesssim m_a\lesssim 10^{-3}\,{\rm eV}$ \cite{KC}. Although very light, axions are legitimate candidates to constitute CDM, since they could have been produced non-thermally in the early universe. Furthermore, they satisfy the two criteria necessary for CDM, {\it i.e.~}(a) they are effectively collisionless (their long range interaction is gravitational) and (b) if their mass is low, a very cold Bose-Einstein condensate of primordial axions could populate the universe today to provide the required dark matter energy density. Indeed, CDM axions have been studied intensely and searches for such a DM particle are underway \cite{ADMX}.

There is compelling evidence that the early universe underwent a period of accelerated expansion, known as ``inflation'' \cite{Guth}. Although several different models for inflation exist \cite{inflation review}, they all share the nontrivial requirement of flatness of the inflaton potential. This in mandatory (a) for inflation to take place and (b) to match the primordial perturbations indicated by the Cosmic Microwave Background (CMB) anisotropies \cite{WMAP}.

One way to obtain a very flat potential is precisely to use a PNGB as the inflaton, as was proposed in natural inflation \cite{Natural, NaturalWMAP}. In this model, the potential of the inflaton field has a particular form, resulting from explicit breaking of a shift symmetry. This symmetry guarantees the flatness of the potential and protects it from too large radiative corrections. Depending on the value of $f_{\rm{infl}}$, the model falls into the large field ($f_{\rm{infl}} > M_{\rm P}$) (where $M_{\rm P}=2.4\times10^{18}$ GeV is the reduced Planck mass) or small field ($f_{\rm infl} < M_{\rm P}$) classification scheme that has been applied to inflationary models. However, the model agrees with the recent observations of the CMB anisotropy by WMAP7+BAO+$H_0$(+SPT) \cite{Keisler:2011aw}\footnote{If data including {\it Clusters} \cite{Keisler:2011aw} are considered, a larger value of $f_{\rm infl}$ is needed.} for $f_{\rm infl} \gtrsim 4\,M_{\rm P}$. In principle it is unclear whether such a high value is compatible with an effective field theoretical description, and whether the global axionic symmetry survives quantum gravity effects. A proposal serving as a solution to this issue was made in ref.~\cite{KNP}, where two axion fields with an effective large decay constant were considered\footnote{A convincing embedding of this field theory proposal into string theory still needs to be found. For a recent discussion see ref.~\cite{Conlon1}.}. In this model, one combination of the fields is massive while the other turns out to be extremely light due to a slight breaking of the symmetry between the couplings of the axions,  moreover giving rise to a large effective decay constant, as required.

While the presently observed acceleration of the universe can be accounted for by a tiny cosmological constant corresponding to an equation of state\footnote{The equation of state for the cosmological matter is $p=w\rho$ where $\rho$ is the energy density and $p$ is its pressure.} of $w = -1$, other equations of state with $-1<w<-1/3$ are not yet excluded by observations (see \cite{Qreview} for a recent review). Indeed, another possibility for dark energy is quintessence \cite{CDS}, where the vacuum energy is not constant but instead  relaxes slowly due to the evolution of a scalar field ({\it e.g.~}quintessence). Moreover, the current vacuum energy density of the classical quintessential field must be of order $E_{\rm vac} \sim 0.003$ eV. This requires the mass of the quintessence to be extremely small, $m_{\rm Q}\sim 10^{-32}$ eV. Such a light scalar field  has the generic problem of driving long-range fifth forces, which are at odds with observations. The standard solution to this problem involves PNGBs with shift symmetries, which forbid higher dimensional operators, rendering the quintessence field stable against radiative corrections. Thus axion fields turn out to be  natural candidates to explain the present day acceleration of the universe too.

In \cite{KN1, KN2} a quintessential axion model was proposed, which also accounts for the QCD invisible axion. The model consists of two axion-like fields where, similarly to \cite{KNP}, the light field is responsible for the present day acceleration, while the heavy field could be identified with the invisible QCD axion. 

According to the above discussion, a natural explanation for the origin of all three basic ingredients of the $\Lambda$CDM model can be found in terms of axion particles, with the additional bonus of providing a solution to the strong CP problem. In this letter we consider this possibility and show that a minimal number of four axions is enough to fulfill all requirements. We motivate our multiaxion system in the context of string theory, where several axions are present in generic compactifications of the theory. It is known that in string theory the common values for the (individual) axions' decay constants turn out to be $f_a\gtrsim 10^{16}$ GeV \cite{SW}. However, there exist mechanisms within heterotic and type IIB string theories, which allow for much lower values for the decay constants \cite{accions, Conlon2}.

Let us note that our approach is different to that of the axiverse \cite{axiverse} since we do not expect a plenitude of light axions to be around at low energies. Indeed, finding a large number of light axion-like particles from string theory has proved hard to arrange, in spite of the plethora of axions present in the theory. For similar reasons our framework differs also from N-flation \cite{Nflation}, where the ``radial" combination of a very large number of axions plays the role of the inflaton and a large decay constant is achieved as the sum of the large number of independent decay constants of all axion fields. Therefore our approach is rather minimal in the sense that we require a limited number of axions, four, to have the required properties to provide a natural origin for the early and present day acceleration as well as a CDM candidate (which includes a solution to the strong CP problem). This minimality might turn out to be a virtue  from the point of view of quantum gravity too, since it might assist in meeting the challenging constraints for multi-field inflationary models described in \cite{Conlon1}.

\section{Potentials for axion cosmology}

Let us begin by briefly reviewing some applications of axion physics in cosmology, including the possibility to describe inflation via axions (natural inflation) or to obtain candidates for dark energy within the quintessence scenario. This will set up the stage in order to investigate richer possibilities, in an attempt to account for all the above cosmological phenomena within a single framework.

\subsection{Two-axion models}

In cosmology nearly flat potentials are of crucial importance. A very attractive way to protect the flatness of a potential is by invoking a shift symmetry. Such symmetries naturally arise in the presence of pseudoscalar fields, such as axions. Then, shallow potentials may be generated by the breaking of such global symmetries.

In single field models, the explicit breaking of a shift symmetry gives rise to a potential of the form
\be \label{1axionpot}
V(\theta)=\L^4\biggl[1-\cos\left(\frac{\theta}{f}\right)\biggl]\, ,
\ee
where $\theta$ is an axion with the shift symmetry $\theta\rightarrow\theta+{\rm constant}$ and $f$ its scale. As it was advocated in \cite{KNP, KN1}, it is often plausible to consider more than one axion in order to meet certain observational constraints. The presence of many generations of axions is not unreasonable in view of the plenitude of global symmetries in string compactifications \cite{axiverse}.
Let us consider the simplest possibility of two axions related to two shift symmetries. The corresponding potential resulting from their breaking is a direct generalization of Eq.~(\ref{1axionpot}) and it is given by
\be \label{2axionpot}
V(\theta,\,\rho)=\L_1^4\biggl[1-\cos\left(\frac{\theta}{f_1}+\frac{\rho}{g_1}\right)\biggl]+\L_2^4\biggl[1-\cos\left(\frac{\theta}{f_2}+\frac{\rho}{g_2}\right)\biggl]\, ,
\ee
where the two axions are denoted as $\theta$ and $\rho$. Clearly, the mass matrix in the $(\theta,\,\rho)$ basis is not diagonal and it is easy to see, {\it e.g.~}by calculating its determinant, that a flat direction exists when the condition
\be\label{2axionflat}
\frac{f_1}{g_1}=\frac{f_2}{g_2}\, ,
\ee
is satisfied. It is then straightforward to determine the physical fields which correspond to linear combinations of $\theta$ and $\rho$. Since in the following we shall perform such an analysis for cases with more axions, here we just state some qualitative features of this procedure. Let us stress that the scales $\Lambda_i$ should exhibit large hierarchies in order to account for the relevant phenomena. This fact obviates the need for exact eigenvectors of the mass matrix (which in some cases are very complicated); indeed, approximate eigenvectors turn out to be enough in our framework. According to the above, assuming without loss of generality that $\L_1 \gg \L_2$, the potential may be written in terms of the physical fields, say $\tilde\theta$ and $\tilde\rho$, as
\be
V(\tilde\theta,\,\tilde\rho)=\L_1^4\biggl[1-\cos\left(\dfrac{\tilde\theta}{f_{\tilde\theta}}\right)\biggl]+\L_2^4\biggl[1-\cos\left(c(f_i,\,g_i)\,{\tilde\theta}+\dfrac{\tilde\rho}{f_{\tilde\rho}}\right)\biggl]\, .
\ee
This form of the potential decouples the physical fields, in the sense that $\tilde\theta$ is a heavy  field, while $\tilde\rho$ is  light when the symmetry condition (\ref{2axionflat}) is nearly satisfied and therefore it has a  high effective scale $f_{\tilde\rho}$. This feature may be utilized in cosmological applications. Indeed, in \cite{KNP} the above mechanism was invoked to account for inflation, identifying the light axion with the inflaton. The increase in the effective axion scale of the physical field $\tilde\rho$ as compared to the scales of the initial fields assisted in meeting the observational bound $f\gtrsim 4\,M_{\rm{P}}$ \cite{NaturalWMAP}. Furthermore, in \cite{KN1, KN2} a different application of the above procedure led to the identification of the two axions as a candidate for quintessence (termed quintaxion) and a QCD axion dark matter candidate.

According to the above it is reasonable to pose the question whether it is possible to use such mechanisms to account for inflation, dark matter and quintessence at the same time. Evidently, an affirmative answer to this question would involve at least three axions. Therefore we begin our pursuit for an answer with the minimal set up of exactly three axions. As we shall see, this set up will prove inadequate to accommodate the QCD axion, however it will pave the road to overcome the difficulties by considering a fourth axion. The latter possibility is investigated in the next section.

\subsection{Three-axion model}

Let us consider three axion fields denoted as $\theta,\,\rho$ and $a$. The first two are considered to be hidden sector axions, while the latter one is the model-independent axion present in all string compactifications. According to the previous discussion a potential for the above fields, generated by the breaking of the associated shift symmetries, has the following form
\be\label{pot}
V(\theta,\,\rho,\,a)=\sum_{i=1}^3 V_i= \sum_{i=1}^3\L_i^4\biggl[1-\cos\biggl(\frac{\theta}{f_i}+\frac{\rho}{g_i}+\frac{a}{h_i}\biggl)\biggl]\, .
\ee
Let us first determine some general properties of this potential. First of all, since the axion $a$ is the model-independent one, it couples universally and therefore it is natural to set $h_1=h_2=h_3\equiv h$. Secondly, expanding the potential around the minimum $\theta=\rho=a=0$ we obtain the mass matrix in this basis
\be\label{pmatrix}
M^2=\begin{pmatrix}
    \frac{\L_1^4}{f_1^2}+\frac{\L_2^4}{f_2^2}+\frac{\L_3^4}{f_3^2}   &   \frac{\L_1^4}{f_1g_1}+\frac{\L_2^4}{f_2g_2}+\frac{\L_3^4}{f_3g_3} &
    \frac{1}{h}(\frac{\L_1^4}{f_1}+\frac{\L_2^4}{f_2}+\frac{\L_3^4}{f_3}) \\ \frac{\L_1^4}{f_1g_1}+\frac{\L_2^4}{f_2g_2}+\frac{\L_3^4}{f_3g_3} & \frac{\L_1^4}{g_1^2}+\frac{\L_2^4}{g_2^2}+\frac{\L_3^4}{g_3^2}   &   \frac{1}{h}(\frac{\L_1^4}{g_1}+\frac{\L_2^4}{g_2}+\frac{\L_3^4}{g_3}) \\
    \frac{1}{h}(\frac{\L_1^4}{f_1}+\frac{\L_2^4}{f_2}+\frac{\L_3^4}{f_3})  &  \frac{1}{h}(\frac{\L_1^4}{g_1}+\frac{\L_2^4}{g_2}+\frac{\L_3^4}{g_3}) &
    \frac{\L_1^4+\L_2^4+\L_3^4}{h^2}
    \end{pmatrix}\, .
\ee
The determinant of this mass matrix is obtained as
\be \label{detm2}
\det M^2=\frac{\L_1^4\L_2^4\L_3^4}{(f_1f_2f_3g_1g_2g_3h)^2}\biggl[f_1f_2g_3(g_1-g_2)+f_2f_3g_1(g_2-g_3)+f_1f_3g_2(g_3-g_1)\biggl]^2\, .
\ee
It may be easily verified that for $f_1=f_2=f_3$ or $g_1=g_2=g_3$, flat directions are obtained. In fact, the general conditions for the existence of exactly flat directions read as
\be \label{flat3}
\frac{f_1}{g_1}=\frac{f_2}{g_2}=\frac{f_3}{g_3}\, ,
\ee
directly generalizing the corresponding result (\ref{2axionflat}) of the two-axion case.

The procedure one may follow in order to determine the physical fields in the present case is straightforward. One has to determine the eigenvalues of the mass matrix, whose eigenvectors will be the physical fields and rewrite the potential in terms of these eigenvectors.

Although the above procedure is straightforward, it becomes increasingly complicated when three or four axions are present, which are the cases we study in the present paper. However, the hierarchy among the scales $\Lambda_i$ facilitates our analysis, since one may instead determine approximate eigenvectors of the potential by applying certain manipulations to the potential itself as we now explain. In order to make the procedure more illuminating we assume at this stage that $f_1=f_3$ and $g_2=g_3$. These choices simplify the determinant of the mass matrix to
\be \label{detm2simple}
\det M^2=\frac{\L_1^4\L_2^4\L_3^4}{(f_1f_2g_1g_2h)^2}(f_1-f_2)^2(g_1-g_2)^2\, ,
\ee
while the conditions (\ref{flat3}) become $f_1=f_2$ and $g_1=g_2$. Finally, as in the previous case, we assume, without loss of generality, that $\L_1\gg\L_2\gg \L_3$.

Let us now define the following new fields:
\bea
&& \varphi = \frac{f_1g_1h}{\sqrt{f_1^2g_1^2+g_1^2h^2+f_1^2h^2}}\biggl(\frac{\theta}{f_1}+\frac{\rho}{g_1}+\frac{a}{h}\biggl)\, , \\
&& \chi= \frac{f_1g_1}{\sqrt{f_1^2+g_1^2}}\biggl(-\frac{\theta}{g_1}+\frac{\rho}{f_1}\biggl)\, , \\
&& \psi= \frac{h f_1}{\sqrt{f_1^2+ h^2}}\biggl(-\frac{\theta}{h}+\frac{a}{f_1}\biggl)\, .
\eea
The field $\varphi$ is the normalized linear combination appearing in the first term of the potential (\ref{pot}). It is an eigenvector up to subdominant $(\L_2/\L_1)^4$ corrections and therefore it is the first physical field. Moreover, it is easy to check that while $\varphi$ is orthogonal to $\chi$ and $\psi$, the latter are not orthogonal to each other. However we will orthogonalize them at a second step. Let us have a look at the form of the potential in terms of these new fields. The three terms $V_i,\,i=1,\,2,\,3$ are given by
\bea \label{V2}
V_1&=&\L_1^4\biggl[1-\cos\biggl(P_1\varphi\biggl)\biggl]\, ,\\
V_2&=&\L_2^4\biggl[1-\cos\biggl(P_2\varphi + Q_2\chi + R_2\psi\biggl)\biggl]\, ,\\
V_3&=&\L_3^4\biggl[1-\cos\biggl(P_3\varphi + Q_3\chi + R_3\psi\biggl)\biggl]\, ,
\eea
where $P_i,\,Q_i$ and $R_i$ are explicit complicated functions of $f_i,\,g_i$ and $h$, with the property that when $f_1=f_2$ and $g_1 = g_2$ the potential is independent of $\chi$ and $\psi$. However, we still need to decouple these two light fields. This decoupling requires a second step, where we ignore the heavy field $\varphi$ and focus on the two other  light fields, which can be separated into a semi-heavy one and a light one. We thus define two new physical fields in this step:
\bea\label{newfis}
&& \chi_1 = \frac{1}{\sqrt{Q_2^2+R_2^2}}\left[Q_2 \, \chi+R_2\,\psi\right]\, ,\\
&& \psi_1 = \frac{Q_2R_2}{\sqrt{Q_2^2+ R_2^2}}\left[\frac{\chi}{Q_2}-\frac{\psi}{R_2}\right]\, .
\eea
Note that these two fields are orthogonal to each other and to $\varphi$. The potential term $V_1$ remains unchanged, while the other two terms now become
\bea \label{V2new}
V_2&=&\L_2^4\biggl[1-\cos\biggl(P_2\varphi +\tilde Q_2\chi_1\biggl)\biggl]\, ,\\
V_3&=&\L_3^4\biggl[1-\cos\biggl(P_3\varphi + \tilde Q_3\chi_1 + \tilde R_3\psi_1\biggl)\biggl]\, ,
\eea
with $\tilde Q$ and $\tilde R$ some new functions of the parameters $f_i,\,g_i$ and $h$. With this potential we have separated the three fields in a way that when either $f_1=f_2$ or $g_1=g_2$ the potential depends only on $\varphi$ and $\chi_1$, so there is one flat direction as expected. If we consider both conditions simultaneously the potential depends only on $\varphi$.
According to the above, we have a clear separation among the three fields, in a heavy, a semi-heavy (or semi-light) and a light field. We can confirm this by looking at the masses and decay constants of the physical fields. Indeed, the masses turn out to be
\bea
m^2_{\varphi}&\simeq&\left(\frac{1}{f_1^2}+\frac{1}{g_1^2}+\frac{1}{h^2}\right)\L_1^4\, ,\nn\\
m^2_{\chi_1}&\simeq&\frac{A}{f_2^2g_2^2(g_1^2 h^2+f_1^2(g_1^2+h^2))^2}~\L_2^4\, , \nn\\
m^2_{\psi_1}&\simeq&\frac{(f_1-f_2)^2(g_2-g_1)^2(f_1^2+g_1^2)(f_1^2+h^2)}{{A}}\L_3^4\, ,\nn
\eea
where
\be
A=(g_1^2 g_2(f_2-f_1)+f_1^2f_2(g_2-g_1))^2(f_1^2+h^2)+(f_1^2+g_1^2)(f_1^2 f_2(g_1-g_2)+(f_2g_1-f_1g_2)h^2)^2\, .\nn
\ee
The decay constants for the axions that appear above are given by (we adopt here the notation $f_x$ for the decay constant of any field $x$)
\bea\label{DC1}
f_{\varphi}&=&\frac{f_1g_1h}{\sqrt{g_1^2 h^2+f_1^2(g_1^2+h^2)}}\, ,\nn\\
f_{\chi_1} &=&\frac{f_2g_2(g_1^2h^2+f_1^2(g_1^2+h^2))}{\sqrt{A}}\, ,\nn\\
f_{\psi_1} &=&\frac{\sqrt{A}}{|f_1-f_2||g_2-g_1|\sqrt{f_1^2+g_1^2}\sqrt{f_1^2+h^2}}\, .
\eea
Having at hand the masses and decay constants of the physical fields in terms of the parameters appearing in the scalar potential, let us now discuss their possible interpretation. First of all, it is clear that nearly flat potentials may be assigned to the fields $\chi_1$ and $\psi_1$. On the other hand, there is one more field left, the $\varphi$, which definitely cannot be related to the QCD axion, since the scale $\Lambda_1$ is the largest one in the model.

According to the above, and keeping in mind that $\L_1\gg\L_2\gg \L_3$, the only reasonable possibility is to consider $\varphi$ to be a heavy auxiliary axion, while attempting to identify $\chi_1$ with the inflaton and $\psi_1$ with the quintaxion. Whether such an interpretation is plausible depends on the possibility to meet the bound
\be\label{inflationbound}
f_{\rm{infl}}\gtrsim 4\,M_{\rm P}\, ,
\ee
for the inflaton as well as the condition on the quintaxion
\be \label{quintbound}
f_{\rm QA}\simeq M_{\rm{P}}\, ,
\ee
for appropriate scales $\Lambda_i$.

There are indeed sets of values for the parameters $f_i$ and $g_i$ which satisfy the above requirements. Indicatively, we suggest the following possible set of values:
\begin{center}
 \begin{tabular}{c*{1}{c}r}
  Parameter   &   Value   \\\hline
  $h$         &   $M_{\rm{P}}$         \\
  $f_1=f_3$   &   0.15\,$M_{\rm{P}}$   \\
  $g_2=g_3$   &   0.125\,$M_{\rm{P}}$  \\
  $f_2$       &   0.125\,$M_{\rm{P}}$  \\
  $g_1$       &   0.15\,$M_{\rm{P}}$   \\
 \end{tabular}
\end{center}
Here we assigned to $h$ the value $M_{\rm{P}}$, which is expected for the model-independent axion in string compactifications. Moreover, all the scales $f_i$ and $g_i$ are subplanckian as they should.
Substituting these values into Eq.~(\ref{DC1}) we obtain the following results:
\bea
f_{\varphi}&\simeq& 0.1\,M_{\rm P}\, ,\nn\\
f_{\chi_1} &\simeq& 5.0\,M_{\rm P}\, ,\nn\\
f_{\psi_1} &\simeq& 1.1\,M_{\rm P}\, ,
\eea
which is indeed in the desired range\footnote{It goes without saying that the above set of values is not the only possible one. A rather wide range in the parameter space may be obtained assuming that $f_1=g_1$ and $f_2=g_2$. For example, requiring that $f_{\varphi} \in (0.1,\,1.0)M_{\rm{P}},\,f_{\chi_1}\in (4.0,\,7.5)M_{\rm{P}}$ and $f_{\psi_1}\in (0.5,\,1.5)M_{\rm{P}}$, we find that $f_1$ lies within $f_1\in (0.142,\,0.257)M_{\rm{P}}$, while the parameter $f_2$ lies approximately within the interval $f_2\in (\frac{3}{4} f_1,\,f_1).$ Moreover, note that the conditions $f_1=g_1$ and $f_2=g_2$ are not indispensable.}.
Let us stress again that although the initial scales were subplanckian, the effective scale for the field $\chi_1$ is transplanckian.

Let us now turn to the masses of the physical fields for the above parameter values. These, in units of $M_{\rm P}^{-2}$, turn out to be
\bea \label{3axmass}
m^2_{\varphi}&=&10^2\,\L_1^4\, ,\nn\\
m^2_{\chi_1} &=&0.04\,\L_2^4\, ,\nn\\
m^2_{\psi_1} &=&0.83\,\L_3^4\, .
\eea
The above masses depend on the scales $\L_i$ for which we have already assumed the hierarchy $\L_1\gg\L_2\gg\L_3$. It is now time to specify these scales. In particular, the smallest scale $\L_3$, related to quintessence, should be of the order of $0.003$ eV, corresponding to the vacuum energy of the universe today. Then, according to Eq.~(\ref{3axmass}), the mass of the quintaxion is of the order $10^{-32}$ eV, which is an acceptable small value. Furthermore, if we assume that $\L_2$, corresponding to the inflation scale, is at the GUT scale\footnote{This is a reasonable assumption, since natural inflation agrees with the observational data for this scale \cite{NaturalWMAP}.} of $10^{15-16}$ GeV, we obtain an inflaton mass of the order $10^{11-13}$ GeV.

Thus we conclude that a potential of the type (\ref{pot}) involving three axions offers the possibility of accounting for inflation and quintessence in the presence of a heavy spectator field. Although this is already an interesting result we shall not delve into further details. Instead we shall proceed to the treatment of a four-axion case which will prove to be richer and to allow for better interpretations.

\section{Four-axion case}

\subsection{General considerations}

In order to go one step further in our interpretation and try to account also for the QCD axion, let us follow the most straightforward way, which amounts to considering a fourth axion. Denoting the four axions as $\theta,\,\rho,\,\phi$ and $a$, the corresponding potential has the form
\be \label{4axionpot}
V=\sum_{i=1}^4 V_i=\sum_{i=1}^4\L_i^4\biggl[1-\cos\biggl(\frac{\theta}{f_i}+\frac{\rho}{g_i}+\frac{\phi}{h_i}+\frac{a}{h}\biggl)\biggl]\, ,
\ee
where, as before, we consider $a$ to be the model independent axion with universal scale $h$. The necessary procedure in order to determine the physical fields, their decay constants and their masses closely follows the three-axion case. The corresponding manipulations are very tedious but totally straightforward. Here we explain in a qualitative way the procedure and we do not present all the intermediate steps, since they do not add anything substantial to the physics of the problem.

The mass matrix in the present case is a $4\times 4$ one and it is of course non-diagonal. Its determinant may be easily computed but it does not directly acquire an illuminating form. Thus, without loss of generality, we assume that $f_1=f_2=f_4$, $g_2=g_3=g_4$ and $h_1=h_2=h_3$. Then the determinant of the mass matrix is given by
\be
\det M^2=\dfrac{\L_1^4\L_2^4\L_3^4\L_4^4}{(f_1 f_3 g_1 g_2 h_1 h_4 h)^2}{(f_1-f_3)^2 (g_1 - g_2)^2 (h_1 - h_4)^2}\, ,
\ee
which directly shows that the conditions for the three expected flat directions simplify to $f_1=f_3$, $g_1=g_2$ and $h_1=h_4$.

A simple way to unveil the nearly flat directions of the potential in the present case follows a three step orthogonalization. Having started with the fields $(\theta,\,\rho,\,\phi,\,a)$, in a first step we define a set of linear combinations of them, say $(\theta^{(1)},\,\rho^{(1)},\,\phi^{(1)},\,a^{(1)})$, such that $\theta^{(1)}$, the argument in the cosine of the first term in the potential (\ref{4axionpot}), is orthogonal to all the rest of the redefined fields. Moreover, similarly to the three field case discussed in section~3, the four terms in the scalar potential are such that the first one $V_1$, depends only on $\theta^{(1)}$, while the other three depend on all the other fields.

In the second step, we leave the field $\theta^{(1)}$ (and therefore $V_1$) unchanged and we construct the fields $\rho^{(2)},\,\phi^{(2)}$ and $a^{(2)}$, which are linear combinations of $\rho^{(1)},\,\phi^{(1)}$ and $a^{(1)}$ such that $\rho^{(2)}$ is orthogonal to $\phi^{(2)},\,a^{(2)}$ and now the second term in the potential $V_2$, depends only on $\theta^{(1)}$ and $\rho^{(2)}$, while the others depend on all fields.

In the third and final step the fields $\theta^{(1)}$ and $\rho^{(2)}$ (and therefore the terms $V_1$ and $V_2$) are left untouched, while out of $\phi^{(2)}$ and $a^{(2)}$ we construct linear combination of $\phi^{(3)}$ and $a^{(3)}$ such that $\phi^{(3)}$ is orthogonal to $a^{(3)}$ and the potential terms are such that $V_3$ depends on the three fields $\theta^{(1)},\,\rho^{(2)},\,\phi^{(3)}$, while the last term $V_4$ depends on all the fields.

Thus at the end of the day, collecting the various terms and adopting the lighter notation $(\theta^{(1)},\,\rho^{(2)},\,\phi^{(3)},\,a^{(3)})\rightarrow (\varphi,\,\chi,\,\psi,\,\omega)$ for the physical fields, the potentials take the form:
\bea \label{4axionpotphys}
V_1&=&\L_1^4\biggl[1-\cos\biggl(A_1\varphi\biggl)\biggl]\, ,\nn\\
V_2&=&\L_2^4\biggl[1-\cos\biggl(A_2\varphi + B_2\chi \biggl)\biggl]\, ,\nn\\
V_3&=&\L_3^4\biggl[1-\cos\biggl(A_3\varphi + B_3\chi + C_3\psi\biggl)\biggl]\, ,\nn\\
V_4&=&\L_4^4\biggl[1-\cos\biggl(A_4\varphi + B_4\chi + C_4\psi+D_4\omega\biggl)\biggl]\, ,
\eea
for some specific constants $A_i,\,B_i,\,C_i$ and $D_i$, being functions of $f_i,\,g_i,\,h_i$ and $h$. We also assume the hierarchy $\L_1\gg\L_2\gg\L_3\gg\L_4$. After the above manipulations the resulting fields are all orthogonal to each other and their potential has the illustrative form (\ref{4axionpotphys}). Moreover, under the three conditions $f_1=f_3$, $g_1=g_2$ and $h_1=h_4$ it depends only on the heavy axion $\varphi$\footnote{Indeed, we have checked that the constants $B,\,C$ and $D$ vanish under the aforementioned conditions.}.

The decay constants and the masses of the physical fields are functions of the parameters $f_i,\,g_i,\,h_i$ and $h$. They are easily determined, however they are rather cumbersome and we shall not present them explicitly here. Instead we shall calculate them below for some specific values of the parameters.

\subsection{Values of decay constants, scales and masses}

Since the general formulae in the four-axion case are complicated, it is far more illustrative to check whether appropriate numerical values for the parameters are adequate to result in reasonable values for the decay constants and the masses of the physical fields. In the present case we would like to exhibit a set of values which may lead to the identification of $\chi$ as the inflaton, $\psi$ as the QCD axion and $\omega$ as the quintaxion. As before, $\varphi$ is a heavy auxiliary field.

Let us choose the following value set of subplanckian scales\footnote{Just as before, other values of the parameters around those presented here are possible and can yield the desired results.}

\begin{center}
 \begin{tabular}{c*{1}{c}r}
Parameter        &   Value                     \\ \hline
$h$              &   $M_{\rm{P}}$              \\
$f_1=f_2=f_4$    &   0.75\,$M_{\rm{P}}$        \\
$g_2=g_3=g_4$    &   0.66\,$M_{\rm{P}}$        \\
$h_1=h_2=h_3$    &   0.75\,$M_{\rm{P}}$        \\
$f_3$            &   0.0000003\,$M_{\rm{P}}$   \\
$g_1$            &   0.75\,$M_{\rm{P}}$        \\
$h_4$            &   0.40\,$M_{\rm{P}}$        \\
\end{tabular}
\end{center}
Substituting these values in the expressions for the decay constants we obtain
\bea
f_{\varphi} &\simeq& 0.4\,M_{\rm{P}}\, ,\\
f_{\chi}    &\simeq& 4.9\,M_{\rm{P}}\, ,\\
f_{\psi}    &\simeq& 0.5\times 10^{-6}\,M_{\rm{P}}\, ,\\
f_{\omega}  &\simeq& 1.0\,M_{\rm{P}}\, .
\eea
It is directly observed that the resulting values are in principle appropriate to describe an auxiliary axion $\varphi$, an inflaton $\chi$, a QCD axion $\psi$ and a quintaxion $\omega$. Indeed the bounds (\ref{inflationbound}) and (\ref{quintbound}) are met, as well as the QCD axion bound
\be \label{qcdbound}
10^9\,{\rm GeV}\lesssim f_{a} \lesssim 10^{12}\,{\rm GeV} \quad \Leftrightarrow \quad 10^{-9}\,{M_{\rm{P}}}\lesssim f_{a} \lesssim 10^{-6}\,{M_{\rm{P}}}\, .
\ee
We note once more that the effective scale of the field $\chi$ is transplanckian although the initial scales were not. This is indispensable for its interpretation as an inflaton.

Whether the above interpretation is really true depends also on the corresponding energy scales. We have already assumed the hierarchy $\L_1 \gg \L_2 \gg \L_3 \gg\L_4$. Let us now specify the scales $\L_i$. The smallest scale $\L_4$ should be of the order of $0.003$ eV, since it is related to the vacuum energy. Moreover, the next to smallest scale $\L_3$ should be the QCD scale ($\sim$ 200 MeV). Finally, $\L_2$ is identified with the inflation scale, which we assume that it corresponds to the GUT scale of $10^{15-16}$ GeV, while the largest one, $\L_1$, is taken to be the Planck scale. These identifications are in accord with the interpretation of the axions which we suggested above. Before discussing further their cosmological evolution, let us also compute their masses. These turn out to be of the following orders
\bea
m_{\varphi} &\simeq 10^{18}\,{\rm GeV}\, ,\\
m_{\chi}    &\simeq 10^{12}\,{\rm GeV}\, ,\\
m_{\psi}    &\simeq 10^{-4}\,{\rm eV}\, ,\\
m_{\omega}  &\simeq 10^{-32}\,{\rm eV}\, .
\eea
We observe the desired hierarchy between the masses. The quintaxion has an extremely small mass of the order of $10^{-32}$ eV, while the QCD axion mass of order $10^{-4}$ eV lies within its acceptable window. Finally, the inflaton mass is of the order of $10^{12}$ GeV.

\section{Interpretation}

Let us now  interpret the cosmological evolution of the four fields of section~4. We divide our analysis in four stages, each for one of the hierarchical scales $\L_i$:
\begin{itemize}
\item Stage 1: The first potential term $V_1$ in Eq.~(\ref{4axionpotphys}) dominates. At this stage, the field $\varphi$ evolves under the effect of this term, since the other three terms are negligible, oscillating around the minimum $\varphi=0$. The fields $\chi,\,\psi$ and $\omega$ remain frozen at their initial value.

\item Stage 2: The second potential term $V_2$ in Eq.~(\ref{4axionpotphys}) dominates. This is the stage of inflation. The field $\varphi$ has a large mass and it has settled to the minimum $\varphi=0$. Moreover, the fields $\psi$ and $\omega$ remain frozen again since the potential does not depend on them at this stage. The field $\chi$ evolves (slow-rolls) under the effect of $V_2$.

\item Stage 3: At later times, the third potential term in Eq.~(\ref{4axionpotphys}) dominates. Apart from the field $\varphi$, which has a large mass and it sits at its minimum, now also the inflaton $\chi$ has rolled down and inflation has ended. The field $\omega$ remains frozen, while $\psi$ now evolves under the effect of $V_3$.

\item Stage 4: At this final stage, the fourth potential term $V_4$ in Eq.~(\ref{4axionpotphys}) now plays a role. The first three fields are very massive compared to $\omega$, which evolves slowly since the decay constant of this field is near the Planck scale, as required for quintessence.
\end{itemize}

\subsection{Axionic see-saw}

\begin{figure}[t]
\centering
\includegraphics[width=1.26\textwidth]{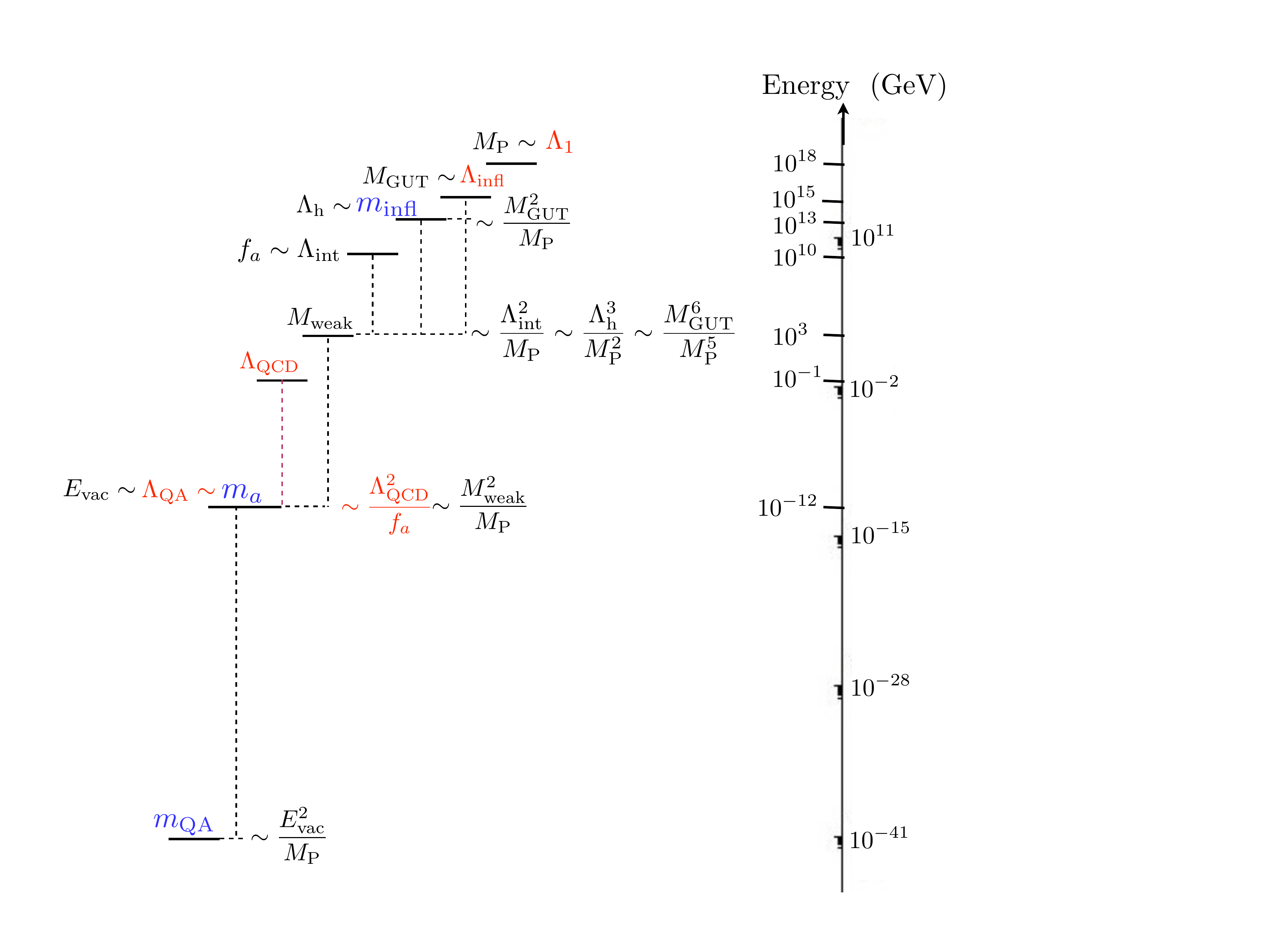}
\caption{Scales, masses and the axionic see-saw.}
\label{fig:scales}
\end{figure}

Let us now discuss some interesting relations among the scales which are relevant in the present framework. A diagrammatic companion illustrating the ensuing discussion appears in Fig.~\ref{fig:scales}. We start with the following benchmark relation between the Planck scale $M_{\rm{P}}$, the weak scale $M_{\rm{weak}}$, the vacuum energy $E_{\rm{vac}}$ and the mass of the quintaxion $m_{\rm{QA}}$:
\be \label{benchmark}
E_{\rm{vac}} \sim \frac{M_{\rm{weak}}^2}{M_{\rm{P}}}\, , \quad m_{\rm{QA}} \sim \frac{E_{\rm{vac}}^2}{M_{\rm{P}}}\, .
\ee
The weak scale is $\mathcal O(\rm TeV)$, while as we have already mentioned, $E_{\rm{vac}}\sim 0.003$ eV and $m_{\rm QA} \sim 10^{-32}\,{\rm eV}$. Moreover, we invoke the hidden sector of heterotic string compactifications in order to relate the weak scale to higher energy scales. Indeed, a dynamical mechanism such as hidden sector gaugino condensation may account for supersymmetry breaking with a gravitino mass at the (multi) TeV scale. In that case the corresponding relation between the scales is \be \label{gc}
M_{\rm weak} \sim \frac{\L_{\rm{h}}^3}{M_{\rm{P}}^2}\, ,
\ee
where $\L_{\rm{h}} \sim 10^{13}\,{\rm GeV}$ is the hidden sector scale, obtained as
\be
\L_{\rm{h}} \sim \frac{M_{\rm{GUT}}^2}{M_{\rm{P}}}\, .
\ee
In the framework we present here, there are four scales $\L_i,\,i=1,\,\dots,\,4$ which are hierarchical as $\L_1\gg\L_2\gg\L_3\gg\L_4$. The highest scale is naturally associated with the Planck scale, {\it i.e.~}$\L_1 \sim M_{\rm{P}}$. Moreover, the second highest scale, the scale of inflation, is identified with the GUT scale, {\it i.e.~}$\L_2 \sim M_{\rm{GUT}} \sim 10^{15-16}$ GeV. We can directly state that the inflaton mass is given by
\be
m_{\rm{infl}} \sim \frac{\L_2^2}{M_{\rm P}} \sim 10^{12-14}\,{\rm GeV}\, .
\ee
Note that this mass is at the intermediate energy scale of the hidden sector, {\it i.e.~}$m_{\rm infl} \sim \L_{\rm h}$. Moreover, the smallest scale is identified with the vacuum energy, {\it i.e.~}$\L_4 \sim E_{\rm vac}$, which is related to the other scales via the relations~(\ref{benchmark}). Finally, the remaining scale is identified with the QCD scale, $\L_3 \sim \L_{\rm{QCD}}$, while the corresponding axion has a mass of the order
\be
m_{a} \sim \frac{\L_{\rm{QCD}}^2}{f_{a}}\sim E_{\rm{vac}}\, ,
\ee
with an invisible axion scale $f_{a}$ of order $10^{10-11}$ GeV. The latter scale, say $\L_{\rm{int}}$, can be also related to the rest as follows. Having in mind the relation~(\ref{gc}) from the gaugino condensation, we may write
\be
M_{\rm{weak}} \sim \frac{\L_{\rm{int}}^2}{M_{\rm{P}}}\, ,
\ee
which entangles the scales $M_{\rm{P}},\L_{\rm{h}}$ and $\L_{\rm{int}}$ as
\be 
\L_{\rm{h}}^3 \sim M_{\rm{P}} \L_{\rm{int}}^2\, .
\ee
Let us stress that such a small value for the decay constant $f_{a}$ (see also the value of $f_3$ in our choice of parameters in section~4) is not trivially achieved in a string-theoretical framework, where the decay constants of model dependent axions are generically higher than $10^{16}$ GeV.

Fig.~\ref{fig:scales} summarizes what we call the axionic see-saw. Given one ratio $R\sim \Lambda_{\rm h}/M_{\rm P}\sim 10^{-5}$ we can parameterize all relevant axionic scales with this one parameter. This may be illustrated as the following chain:
$$
M_{\rm P}\sim m_{\varphi}\,\overset{R}\longrightarrow\, \L_{\rm h}\sim m_{\rm infl}\, \overset{R^2}\longrightarrow\,M_{\rm weak}\,\overset{R^3}\longrightarrow\, E_{\rm vac}\sim m_{a}\,\overset{R^6}\longrightarrow\, m_{\rm QA}\, .
$$
All the three ``useful'' axions fit into the scheme. The relations could be motivated from diverse UV completions in various string schemes. There are known mechanisms, such as {\it e.g.~}the breaking of accidental symmetries in the heterotic theory \cite{accions} or within the framework of the large volume scenario in type IIB string theory \cite{Conlon2}, which allow for lower values for the decay constants. Thus here we assume that one of such mechanisms operates in our scenario but we leave a more detailed discussion for future work.

\subsection{The role of supersymmetry}

Previous discussions usually assumed the presence of supersymmetry at the weak scale as a solution of the hierarchy problem. But as we have said, axions do not really need susy, although both schemes are compatible with each other. One of the strong arguments for weak scale susy, the existence of a WIMP dark matter candidate is no longer valid in the axion framework. So we do not need to be prejudiced about the value of the susy breakdown scale. Fine tuning questions could be relevant towards low values of $M_{\rm susy}$. They might be relevant for the stability of the weak scale but the smallness of the vacuum energy cannot be explained in that way. So we would need alternative arguments which we do not discuss here in detail.

Given this situation we would now have to see how the susy scales like the gravitino mass, the soft breaking terms of the supersymmetric partners in the MSSM and the $\mu$-parameter
fit into the axionic see-saw scheme of Fig.~\ref{fig:scales}. As we have explained above, in previous works \cite{KN2, Lowen:2008xh} some identifications have been made, but they were usually done with the prejudice for weak scale susy.

So let us here be more general and discuss the framework from a top-down approach in string theory. As we have seen, string theory provides its intrinsic mass scale $M_{\rm string}$ that is relevant for the axion scheme. Typically all axion decay constants are of order of $M_{\rm{string}}$ and the small size of the scale of the QCD axion
is a challenge in string model building \cite{accions, Conlon2}.

Could there be a relation between the axion scale and the scale of susy breakdown? As we already remarked, consistent string theories in $D=10$ require supersymmetry (most notably due to the requirement of absence of tachyons). Of course supersymmetry is broken at some scale, but that could be a scale sufficiently low compared to $M_{\rm string}$. We could argue that this scale could be identified with the scale $f_a\sim 10^{9} - 10^{12}$ GeV of the QCD axion. This leaves open the two especially interesting possibilities: weak scale susy or as an alternative susy at the scale of the invisible axion.

\subsection{The susy scale and hints from the LHC}

LHC is the only present experiment that can provide hints about the fate of supersymmetry. So let us now have a look at the preliminary LHC data and see whether
we can learn something.

\begin{itemize}
\item No signs of susy have been seen yet. This implies at least (multi) TeV range for the supersymmetric partners in the MSSM.
\item There is now strong evidence for a Higgs boson with a mass around 125 GeV. This is

\begin{itemize}

\item rather high for the MSSM (heavy susy particles are needed),

\item but rather low for the SM. The quadlinear Higgs self coupling $\lambda$ runs to zero at a scale smaller than the Planck scale (for a recent calculation see ref.~\cite{Degrassi:2012ry}). This is the scale where new physics is needed to complete the SM.

\end{itemize}

\item Interestingly enough, this scale coincides with the axion scale of the QCD axion.

\item We could now speculate that the susy scale is identified with the scale of the axion decay constant, describing a remote supersymmetry (tele-susy). Special properties of
the Higgs fields, for example in the models of the MiniLandscape \cite{Krippendorf:2012ir}, provide shift symmetries within extended susy that lead to a vanishing $\lambda$ at the scale of supersymmetry breakdown.

\end{itemize}

So we could thus consider the SM without (weak scale) susy and extrapolate to higher energies till we reach the region where $\lambda$ turns negative. We do not need a WIMP candidate for CDM matter as we will use the axion and we postulate the axion scale as the scale of broken susy \cite{Hebecker:2012qp, Ibanez:2012zg}. Of course, this ``tele-susy'' will no longer be able to solve the hierarchy problem of the weak scale. In addition we have to consider the fine tuning problem of the quintessential axion (that would not be solved by weak scale susy either). Thus we have to look for alternatives to solve these hierarchy (or fine-tuning) problems of nature, such as {\it e.g.~}anthropic reasonings \cite{Susskind}.

Now we can ask the question how this speculation about tele-susy fits in the axionic see-saw scheme of Fig.~\ref{fig:scales} and how it differs from previous considerations that assumed weak scale susy. First of all, this question partially regards the value of the gravitino mass and of the $\mu$ term in this scheme. As for the $\mu$ term, note that it is a supersymmetric mass term that could decouple from the susy scale. Even in the case of tele-susy we could have a $\mu$ term of the size of the weak scale and provide a small tree-level mass for the Higgs boson, even if the gravitino mass is identified with the axion decay constant.

In an alternative scenario the vacuum energy might be somehow related to the gravitino mass, maybe via a volume suppression or a suppression with the mechanism of a doubly suppressed gravitino mass \cite{Lowen:2008xh}. Further relevant questions concern the preferred schemes of susy breakdown and mediation, as well as the relation between soft mass terms, gravitino mass and vacuum energy. Of course, one of the central points of tele-susy regards the origin of the susy breakdown scale and its coincidence with the scale of the QCD axion. It might find a solution within the axionic see-saw in very similar way as the $\mu$ term in the case of weak scale susy \cite{KNmu}. A careful discussion of these questions will be the subject of future research.

\section{Conclusions}

The announcement of the Higgs particle exhibited in the most explicit way that we are in the middle of a discovery era in particle physics. Furthermore, promising experiments and 
observations in the cosmological front are underway. However, particle physics and cosmology involve a number of well separated scales which often make it difficult to discuss their phenomena simultaneously. In the present paper we presented a pattern of scales which brings them closer: the axionic see-saw.

The axionic see-saw is a pattern of energy scales which emerges out of the impact between certain aspects of particle physics beyond the standard model and cosmology. On the one hand it is related to an effective field theory of ``useful'' axions. The latter are axions which play an important role in discussions of three key ingredients of cosmology, namely inflation, dark matter and dark energy. In the present paper, we examined to what extent it is possible to obtain axion candidates for the inflaton and quintessence as well as for cold dark matter via axions. The result is that starting with four axions and a potential arising from the breaking of their corresponding shift symmetries, it is indeed possible to achieve the above picture. On the other hand, it is meaningful to ask how supersymmetry and its relevant scales fit into this axionic see-saw pattern. We discussed this question under the light of the implications of the current experimental data and without prejudice on the value of the supersymmetry breaking scale. One possibility is of course weak scale supersymmetry, which has been discussed before. A novel proposal is motivated by the observation that a Higgs boson mass of 125 GeV implies the vanishing of its self coupling at a scale which approximately coincides with the scale of the QCD axion. This observation allows us to speculate that the new physics which should arise at that intermediate scale in order to complete the standard model could be a remote supersymmetry which we called tele-susy.

Tele-susy is further motivated by the fact that the most promising candidate for an ultraviolet completion of the standard model, string theory, requires supersymmetry for its internal consistency. Therefore, even if the LHC disfavors weak scale supersymmetry in the future, a tele-susy has to operate at an intermediate scale between the weak and the Planck scale. The existence of a dark matter candidate, a strong motivation for weak scale supersymmetry, is not lost in this scheme, this role being played by the QCD axion. Regarding the gauge hierarchy problem one has to reside to anthropic arguments for the moment. Needless to say that in a unified picture, such as the one we discussed in this paper,
there is anyhow a far more severe hierarchy related to the observed vacuum energy of the universe.

An interesting direction for future investigations regards possible realizations of the above picture in concrete models. In a top-down approach one should look for string compactifications which could provide the useful axions discussed above. One issue that has to be addressed explicitly is the low scale of the QCD axion compared to the Planck scale, which is a challenge in string model building, that might be explained via the scale of supersymmetry breakdown. In addition, in string compactifications where a plethora of axions are present it would be interesting to explore how the three ``useful'' axions are singled out.

\vspace{10pt}

\section*{Acknowledgments}

A.C. is grateful for the hospitality during a visit to the University of Groningen, where part of this work was done. This work was partially supported by the SFB-Transregio TR33
``The Dark Universe" (Deutsche Forschungsgemeinschaft) and the European Union 7th network program ``Unification in the LHC era" (PITN-GA-2009-237920). E.E. thanks the Bonn--Cologne Graduate School for support.

\end{document}